\documentclass[12pt]{article}
\begin{document}
\begin{large}
\title{\bf { Could quantum statistical regularities derive from a measure on the boundary conditions of a classical universe?}}
\end{large}
\author{Bruno Galvan \footnote{E-Mail: bgalvan@delta.it}}
\date{June 1998}
\maketitle
\begin{abstract}
The problem of defining the boundary conditions for the universe is considered here in the framework of a classical dynamical theory, pointing out that a measure on boundary conditions must be included in the theory in order to explain the statistical regularities of evolution. It is then suggested that quantum statistical regularities also could derive from this measure\@. An explicit definition of such a measure is proposed, using both a simplified model of the universe based on classical mechanics and the non-relativistic quantum mechanics formalism\@. The peculiarity of such a measure is that it does not apply to the initial conditions of the universe, i.e. to the initial positions and momenta of particles, but to their initial and final positions, from which the path is derived by means of the least action principle\@. This formulation of the problem is crucial and it is supported by the observation that it is incorrect to liken the determination of the boundary conditions of the universe to the preparation of a laboratory system, in which the initial conditions of the system are obviously determined\@. Some possible objections to this theory are then discussed\@. Specifically, the EPR paradox is discussed, and it is explained by showing that in general, a measure on the boundary conditions of the universe generates preinteractive correlations and that in the presence of such correlations Bell's inequality can no longer be proven true\@. Finally, it is shown that if one broadens the dynamical scheme of the theory to encompass phenomena such as particle decay and annihilation, the least action principle allows for an indeterministic evolution of the system.
\end{abstract}

\section{Introduction.}
Attempts to make quantum mechanics a universal theory, i.e. able to globally represent the physical reality of the universe, clash with the problem of wave packed reduction, which some authors consider to be the only genuine paradox of quantum mechanics: when and why does it take place? Another way to formulate this problem may be the following: how is it possible to obtain a classical representation of the evolution of the universe, i.e. one which is consistent with our perception of reality, from the universal wave function? Several theories and interpretations attempt to solve the problem; a brief classification of the most relevant ones is given for instance by Penrose \cite{penrose}. Such theories generally differ in how they answer the following crucial questions: (i) is the wave function a physical entity or rather a mere mathematical tool for calculating the results of measurements? (ii) is the collapse of the wave function a real physical phenomenon, or it is just an apparent one? (iii) is the wave function the most complete representation of physical reality, or are more complete ones possible?

Many of these theories (but not all of them) are non-antagonist, i.e. do not predict experimental results which are new or different from those predicted by ortodox quantum mechanics. Accordingly, in most cases it is not possible to perform experimental tests able to discriminate among them. This is probably one of the reasons why a theory approved unanimously by the scientific community is still apparently not available.

This paper proposes a new non-antagonist theory - which for the sake of convenience will be referred to hereafter as {\it quantum-classical} theory - in which particles are described by phase space paths, as in classical mechanics, and the laws controlling their evolution are the least action principle and a measure on the boundary conditions of the universe. The background principles of this theory are outlined in this introduction.

\vspace{.5cm}
In classical dynamical theories, such as classical mechanics or general relativity, the evolution of a system is determined by the law of motion and by boundary conditions. For instance, in the case of a hamiltonian system of particles, the law of motion is given by Hamilton's equations, while the boundary conditions are the initial positions and momenta of the particles. Boundary conditions of this kind are also called initial conditions. For a hamiltonian system (as for all classical dynamical systems) there is another kind of boundary conditions, namely the set of particle positions at two times $t_1$ and $t_2$; the path of the system between these two times is determined by Hamilton's least action principle, which acts as a dynamical law. The existence of this kind of boundary conditions will be highly relevant for this paper.

The fact that boundary conditions are needed in order to define the evolution is not a problem in the study of open systems, i.e. of systems undergoing a preparation phase enforced by an external system: for such systems the boundary conditions, or rather the initial conditions, are determined by the preparation phase.

Different considerations apply to a closed system, such as the universe, for which there is no external system subjecting it to a preparation phase. Therefore, defining the boundary conditions of the universe poses some relevant questions, such as: (i) is a law for boundary conditions necessary? (ii) if so, to what kind of boundary conditions does such a law apply? To initial conditions, as in a system subjected to preparation, or to the boundary conditions associated with the least action principle? (iii) are there observable regularities of evolution which derive from the law on boundary conditions? (iv) Is it possible to achieve an explicit formulation of this law?

This paper addresses these problems and offers possible answers to them. Specifically: (i) it will be shown that a law for boundary conditions is actually needed in order to explain statistical regularities of evolution and will be expressed here as a measure; (ii) backed by the observation that determining the boundary conditions of the universe is different from preparing a system for a laboratory experiment, this paper will propose that the boundary conditions to which this measure applies should be constituted not by the initial conditions, but rather by the conditions related to the least action principle; this proposal has deep conceptual implications, such as the possible appearance of preinteractive correlations; (iii) this paper will propose that not only ``classical" statistical regularities, such as the statistical regularity of coin tossing results, but also the statistical regularities predicted by quantum mechanics can be derived from this measure; (iv) finally, an explicit definition for this measure will be proposed.

The theory developed in this paper is based upon the dynamical structure of a hamiltonian particle system; accordingly, the quantum phenomena it considers are only those predicted by the Schr\"{o}dinger equation. However, since its foundations - the principle of least action and a measure on the boundary conditions of the universe - are highly general, it does not seem unfeasible to extend this theory to encompass phenomena predicted by quantum field theory, such as particle decay and annihilation. Such an extension is sketched out in section 13 in order to show how the deterministic character of the theory in its hamiltonian formulation is lost when these phenomena are considered.

\section{A simplified model of the universe}

In order to provide an accurate exposition of the quantum-classical theory, a simplified model of the universe, based on classical mechanics, is defined and used. This model does not seek to be a cosmological theory of the universe; therefore no cosmological inferences, for instance about galaxy distribution or the age of the universe, should be made from it.

In this model we picture the universe as a hamiltonian system of $N$ particles with mass $m_1,\ldots,m_N$ interacting through potentials $V_{i,j}({\bf x}_i-{\bf x}_j)$ which decrease sufficiently quickly as the distance between particles tends to infinity. The phase space is $R^{6N}$, i.e. the Cartesian product of the configuration space and the momentum space of the $N$ particles. The Hamiltonian is:
\begin{equation}
	H=\sum_{i=1}^{n} {\frac{{\bf p}_i^2}{2m_i}}+ {\sum_{i<j}V_{i,j}({\bf x}_i-{\bf x}_j)}.
\label{1}
\end{equation}
The law of motion is defined by Hamilton's equations. A path is a curve $\gamma:R\rightarrow  R^{6N}$ satisfying Hamilton's equations. The position and momentum of the path $\gamma$ at the time $t$ shall be indicated by $\gamma_x(t)$ and $\gamma_p(t)$ respectively. In order to represent an initial big-bang at the time $t=0$, we will consider, only for $t\geq0$, those paths for which at the time $0$ all the particles are located in the space point $x=0$. In the initial part of the discussion it will also be assumed that a final time $T$ exists for the universe. This assumption will be removed later on by taking the limit for $T\rightarrow \infty$.

Briefly, the set of possible paths in the simplified model of the universe, indicated as $\Gamma_T$, consists of the curves $\gamma:[0,T]\rightarrow R^{6N}$ that satisfy Hamilton's equations and are such that $\gamma_x(0)=0$.

\section{The need for a measure on the boundary conditions of the universe.}

Many phenomena in the evolution of our universe appear to be indeterministic. This is considered possible even in a deterministic universe, due to the generally incomplete knowledge of the initial state of the systems being observed. Consider, for instance, a coin-tossing machine, consisting of a mechanism which tosses a coin placed in a suitable slot. Machine preparation consists in loading the tossing mechanism and in placing the coin in the slot. Clearly, if we repeat the toss several times we obtain different results. This is explained by assuming that in the distinct trials there are differences at the molecular level in the tossing mechanism or in the coin or in the air into which the coin is tossed, determining different results.

In several phenomena whose evolution appears to be indeterministic, however, one still finds statistical regularities. Suppose, for instance, that we perform 1000 tosses with the coin-tossing machine. In order to avoid symmetry considerations which could be misleading, assume that one side of the coin is heavier than the other, in such a way that 1000 tosses produce a frequency of heads of, say, 48\%. We cannot yet call this a statistical regularity of evolution, since if we repeat 1000 times an experiment which has two possible outcomes, this will inevitably yield one percentage or another of one of the two results. Nonetheless, we can confirm that this rate actually describes a statistical regularity by repeating the 1000-toss sequence many times and measuring with good approximation the same rate in all the sequences. It is therefore reasonable to ask oneself from what law does this statistical regularity arise.

First of all, let us test whether it can be inferred exclusively from the law of motion. If so, all the boundary conditions of the universe that imply making those 1000 tosses should determine a sequence of outcomes with 48\% heads. It is rather reasonable to assume that this does not occur and that instead there are particular initial conditions which determine a heads rate other than 48\% (this fact will be even more evident in the next example). Therefore, since this statistical regularity also depends on the choice of the boundary conditions of the universe, the law of motion alone is not sufficient to account for it.

In order to explain this statistical regularity we are therefore forced to introduce a new law which must rule the choice of the boundary conditions of the universe, allowing to choose only those boundary conditions leading to 48\% heads in all sequences of multiple tosses of the coin. Following Popper \cite{popper}, we will represent this law by means of a measure on the set of boundary conditions of the universe. This measure acts as a law as follows: it establishes that the heads rate is 48\% assigning an extremely small measure, which is vanishing in the limit, to the subset of the boundary conditions that lead to heads rates different from 48\%.

Let us give a precise mathematical example of this argument. Consider a dynamical theory composed of by two elements: a set $S$, the phase space, and a map $G:S\rightarrow S$, the law of motion. The law of motion associates with a state $s\in S$ the state $G(s)$ in the following instant. A path is a sequence $\{a_n\}$ of elements of $S$ such that $a_{n+1}=G(a_n)$. Once the element $a_0$ of the path, i.e. the boundary condition, have been set, the path is univocally determined. Clearly, this theory consists of the same two basic building blocks of a classical dynamical theory, namely the law of motion and the boundary conditions. A practical example is given by a phase space represented by the interval $[0,1)$ of the real axis and by a law of motion represented by Bernoulli's map $G(x)=2x$ (mod 1). It is useful to understand how Bernoulli's map behaves in terms of the binary expansion of a number: this expansion is, for a number in the interval $[0,1)$, of the form $0.01100101\ldots$, i.e. consists of an integer $0$ and a ``decimal" part composed of an infinite series of $0$'s and $1$'s; Bernoulli's map acts by shifting the binary series one place to the left.

A statistical measurement that can be taken in this universe is, for instance, the frequency with which the state of the universe belongs to the first half of the interval $[0,1)$. More precisely, we will study the limit of $n_0/n$ as $n\rightarrow \infty$, where $n_0$ is the number of instants for which the state belongs to the first half of the interval during the first n instants of the evolution, or equivalently, the number of zeroes present among the n leading digits of the binary series of the boundary condition. Such a limit, if it does exist, will be referenced by $\alpha$.

The law of motion alone does not allow to predict the value of $\alpha$, since there are initial conditions which lead to different values of $\alpha$. Consider for example the two initial conditions $a_0\equiv1/3=0.01010101\ldots$, leading to $\alpha =1/2$, and $a'_0\equiv2/7=0.01001001001\ldots$, which leads to $\alpha=2/3$. There will also be initial conditions for which the limit $n_0/n$ does not exist.

In order to predict the value of $\alpha$, a further element must therefore be added to the theory: a measure on the boundary conditions space, in this case the interval $[0,1)$. Such a measure determines for $\alpha$ a particular value $\alpha_0$ if the measure of the set of initial conditions for which $\alpha\neq\alpha_0$ (or for which the limit $n_0/n$ does not exist) is null. For instance, it can be demonstrated that the Lebesgue measure on the interval $[0,1)$ determines the value $1/2$ for $\alpha$. It is possible to define, on the interval $[0,1)$, measures which determine any value between 0 and 1 for $\alpha$ or determine any kind of statistical regularity for the paths. Reference should be made to the appendix for further details.

\vspace{.5cm}
The impossibility for a purely deterministic theory (i.e. consisting solely of the law of motion) to account for the statistical regularities of evolution has been widely discussed by Popper \cite{popper} and by Land\'{e} \cite{lande}. In particular, the example of the coin-tossing machine and the possibility to explain statistical regularities by means of a measure on the initial conditions were taken from the first author. Popper himself, however, shortly after outlining this solution, criticized it, defining it as an ``irreducible and miraculous statistical distribution of initial conditions". Land\'{e}, too, criticized determinism by saying that ``A distribution of effects satisfying the laws of error theory requires, just from the determinist's viewpoint, a corresponding random distribution of causes at an earlier time and from there to a still earlier time. A program of giving a strictly deterministic theory of statistically distributed events leads nowhere".

We do not share these criticisms. As regards Popper's objection, we observe that normally it is useless to ask oneself why a law exists or where it comes from. This is not done, for instance, for the law of motion, and in our opinion there is no reason to do so for the measure on boundary conditions either. Acknowledging the existence of a law and representing it mathematically in a simple way is sufficient to build a meaningful theory; the validity of the theory will be confirmed or denied subsequently by comparing its predictions with experimental data.
As regards the issue of regression to infinity posed by Land\'{e}, it can be solved, and indeed has been solved, by assuming a cosmological perspective and by speaking of boundary conditions of the universe. Besides, this problem does not arise at all in the example of the universe evolving according to Bernoulli's map.

Let us summarize the arguments carried on so far with the following statement:
\begin{quote}
{\it A classical theory for the universe must include, in addition to the law of motion, a measure on the boundary conditions; such a measure is needed in order to account for statistical regularities of evolution.}
\end{quote}

In the above discussion we examined a ``classical" statistical regularity, such as the heads rate in coin tosses. We will briefly discuss hereafter the possibility that the second law of thermodynamics and the arrow of time may also be statistical phenomena which derive from the measure on the boundary conditions of the universe.

There is an important class of statistical regularities of evolution whose nature is normally considered incompatible with classical dynamical theories, such as classical mechanics or general relativity: it consists of quantum mechanical regularities. This paper will instead examine the possibility that this kind of regularities, too, may be explained by a suitable measure on the set of boundary conditions of a classical universe.

\section{What kind of boundary conditions for the universe?}

In the simplified model of the universe defined in section 2, the initial position $\gamma_x(0)$ of paths is fixed. Accordingly, from a mathematical point of view the possible boundary conditions are the following: initial momentum $\gamma_p(0)$ (which will be referred to hereinafter as type $I$ conditions) and final position $\gamma_x(T)$\footnote{ Actually, defining the initial and final positions does not identify the path univocally; we will not worry about this for the time being but will come back to it later.} (type $F$ conditions). The question is therefore as follows: taking into account that the measure is a law ruling the choice of boundary conditions, to which of these two types of boundary conditions should it be applied? 

It seems obvious that it should be applied to type $I$ conditions, i.e. to initial conditions. These are in fact the boundary conditions that are set in the preparation phase of any laboratory experiment; it makes no sense, from the physical point of view, to think of performing an experiment in which type $F$ boundary conditions are set, i.e. in which the initial and final positions of the system being observed are set. 

However, it is incorrect to liken the preparation of a laboratory system to the determination of  the boundary conditions of the universe. In the first case, the person who determines the initial conditions, i.e. the experimenter, is immersed in the same time flux as the system being observed. In the second case, the question clearly arises of ``who" determines the boundary conditions of the universe and of ``how" this is done. Leaving aside the metaphysical issues, which are entirely beyond the aims of this paper, we assume that in this case the boundary conditions of the universe are determined by an ``entity" on which we will not speculate except to hypothesize that it is {\it external} to the universe, and thus external to its time flux. This means that differently from an experimenter with respect to his laboratory system, such an entity perceives the universe not as a dynamical system, evolving with time, but rather as a static element whose configuration already appears fully defined at all instants. In this framework there it is no paradox in thinking that this entity, in order to determine the path of the universe, could use type $F$ boundary conditions. A discussion about the possibility of looking at the evolution of the universe from an atemporal point ov view may be found in Price \cite{price1}.

Let us examine the two possibilities - type $I$ or type $F$ boundary conditions - in the light of the causality principle, or to be more exact, in the light of the antecedence principle, which assumes that a cause must precede the effect in time (Born \cite{born}). Choosing type I would appear to satisfy the antecedence principle, putting the cause - i.e. the initial momenta of the particles - before the effects - the path of the universe; on the other hand, choosing type $F$ would appear to violate this principle by putting the cause - the final positions of the particles - after the effects. Actually, one cannot say that choosing $I$ satisfies the antecedence principle because, as noted by Born, there is not even a cause-effect relation between the initial conditions and the subsequent path of the universe. Due to the time-symmetry of the law of motion, the state of a classical system at any instant in fact completely determines the state of the system at any other past and future instant; therefore one cannot claim that the state of the system at a given time $t_1$ {\it causes} the state of the system at a given time $t_2>t_1$ any more than one can claim the reverse. This means that selecting type $I$ conditions is equivalent to atemporally selecting the whole path of the universe, i.e. to choosing an element in the space $\Gamma_T$. The same considerations hold for type $F$ conditions. In conclusion, we can say that neither the causality principle nor the antecedence principle are involved in using either kind of boundary conditions.

The reader may question the need for this metaphysical discussion. Besides, as we will see, the problem of applying the measure to one type of boundary conditions or the other is irrelevant from the mathematical point of view, since the measure can be transferred in a natural way from one type to the other. However, this discussion is motivated by the fact that, as we will see, a measure on path space can give rise to highly counterintuitive phenomena such as preinteractive correlations. Stressing the conceptual difference between choosing boundary conditions for the universe and preparing a laboratory system will help in understanding and accepting these phenomena.

We now put forward what is certainly the most important proposal of this paper: namely, that the correct boundary conditions for the universe are type $F$ conditions, i.e. the final positions of the particles in the universe. In the next section we will define on them a measure which will be called {\it quantum measure}.

\section{The quantum measure.}

We shall express the quantum measure by means of a density $\rho_{FQ}(x)$ on the space $R^{3N}$ of final positions, where the index $F$ refers to the type of boundary conditions to which it applies. In order to define $\rho_{FQ}(x)$, assume that the universe is a quantum system; it will therefore be described at each instant $t$ by a wave function $\psi(x,t)\in L^2(R^{3N})$. At the time $t=0$ the particles are concentrated at the point $x=0$; therefore we set $\psi(x,0)=\langle x|0\rangle$, where $|0\rangle$ is the improper eigenstate associated with the eigenvalue 0 of the position operators $Q$, where 
$Q=({\bf Q}_1,...,{\bf Q}_N)=(Q_{1x},Q_{1y},Q_{1z},...,Q_{Nx},Q_{Ny},Q_{Nz})$. The probability density of finding the universe in the position $x$ at the final time $T$ will therefore be $|\langle x|e^{-iHT}|0\rangle|^2$ (where we have set $\hbar =1$). Accordingly, we will define $\rho_{FQ}$ by setting:
\begin{equation}
\rho_{FQ}(x)\equiv |\langle x|e^{-iHT}|0\rangle|^2.
\label{2}
\end{equation}
Therefore, in the quantum-classical theory the universe evolves according to a classical path, i.e. a path which satisfies the least action principle, but the probability that at the final time $T$ it will be in a given position is the probability given by quantum mechanics: quantum mechanics determines the probability of the final position, whilst the least action principle determines the path.

We will now show that the quantum-classical theory can be considered as a way to extract a representation of evolution in classical terms from the wave function of the universe. The ``Consistent histories approach to quantum mechanics" \cite{consistent} attempts to achieve the same result but retains the wave function as a complete description of physical reality.

It is common knowledge that the mathematical formalism of quantum mechanics provides no criteria for determining when collapse of the wave function occurs as a consequence of the measuring process: i.e., whether it occurs when a microscopic system interacts with a macroscopic measuring device or - with reference to the example of Schr\"{o}dinger's cat - when the instrument activates the chemical mechanism that causes the death of the cat, or when the experimenter enters the laboratory to check whether the cat is alive or dead. Taking this argument to its extreme consequences, one can think of the universe as a huge quantum system whose state in every time t is represented by the wave function $\psi (x,t)$. At the final time $T$, an observer external to the universe measures the position of the universe, obtaining a given result x and causing the collapse of the wave function. The problem is then to infer from this result a classical description of the evolution of the universe between the times $0$ and $T$. The most natural inference is obtained by considering a path joining the points $(0,0)$ and $(T,x)$ which minimizes the action, exactly as proposed by the quantum-classical theory.

This argument is based on the following conjecture, which we will call conjecture $Q_1$: {\it all the measurements performed in the universe can be brought back to a position measurement made at the final instant of the universe}. Let us present this conjecture in more detailed terms. Let $\psi (t_I)$ be the wave function of the universe at a given time $t_I$ immediately preceding a measurement which has two possible results, and let $\psi_1(t_F)$ and $\psi_2(t_F)$ be the two wave function of the universe, at the time $t_F$ immediately after the measurement, which correspond to the two results of the measurement. One will obviously have $\psi(t_F)=\psi_1(t_F)+\psi_2(t_F)$. Many authors \cite{feynman}\cite{bell1} acknowledge that if we include the measuring devices in the wave function, then every measurement can be regarded as a position measurement. In our case, this means that the two wave functions $\psi_1(t_F)$ and $\psi_2(t_F)$ are spatially disjoined, i.e. that two disjoined space projectors $E(\Delta_1)$ and $E(\Delta_2)$ exist such that $E(\Delta_1)\psi(t_F)=\psi_1(t_F)$ and $E(\Delta_2)\psi(t_F)=\psi_2(t_F)$. To accept this, consider that the two wave functions are disjoined if they localize even just a single particle of the universe in two different positions; this must certainly occur for $\psi_1(t_F)$ and $\psi_2(t_F)$, since they describe the measuring device that recorded two different results, and therefore is in two macroscopically different configurations: pointers in different positions, ink drops on computer printouts plotting different characters, and so on. Conjecture $Q_1$ proposes to strengthen this property by assuming it to be {\it persistent}, i.e. by assuming that the wave functions $\psi_1(t)$ and $\psi_2(t)$ remain disjoined for every time $t\geq t_F$, and therefore at the final instant of the universe too. We will not discuss conjecture $Q_1$ any further in this quantum formulation, since we will reformulate and discuss it later in classical form.

Let us now leave momentarily aside the discussion of the physical aspects of the quantum-classical theory while we specify its mathematical and formal features in the following sections.

\section{Measures on sets of paths.}

This section will show: (i) that a $\sigma$-algebra and a measure on the boundary conditions of the set of paths $\Gamma_T$ induce a $\sigma$-algebra and a measure on $\Gamma_T$ itself; (ii) that a $\sigma$-algebra and a measure on type $I$ boundary conditions induce a $\sigma$-algebra and a measure on type $F$ boundary conditions, and vice versa; and finally, (iii) that a measure on the set $\Gamma_T$ induces a density $\eta(s,t)$ on phase space which satisfies Liouville's equation.

Let $\Gamma$ be a set; we call {\it parametrization} of $\Gamma$ a map $h:\Gamma\rightarrow R^n$. With reference to the concepts introduced earlier, the set $\Gamma$ corresponds to a set of paths and parametrization is the map that associates a path with its boundary conditions. The set $R^n$ has the $\sigma$-algebra ${\cal B}^n$ of Borel's sets. We indicate by $h^{-1}({\cal B}^n)$ the class of subsets of $\Gamma$ defined as follows: $h^{-1}({\cal B}^n)\equiv \{\Delta\subseteq \Gamma:h(\Delta)\in {\cal B}^n\}$. It is easy to show that $h^{-1}({\cal B}^n)$ is a $\sigma$-algebra, and that it is the minimal $\sigma$-algebra with respect to which the map $h$ is measurable. The individual points $x\in R^n$ are measurable and therefore their counter-images $h^{-1}(x)$ are all and only the atoms of the $\sigma$-algebra $h^{-1}({\cal B}^n)$ \footnote{A set $A$ belonging to a class of sets ${\cal A}$ is an atom for ${\cal A}$ if $A'\in {\cal A}$ and $A'\cap A\neq\emptyset$ implies $A\subseteq A'$.}. Let $h_I:\Gamma\rightarrow I=R^n$ and $h_F:\Gamma\rightarrow F=R^m$ be two parametrizations of $\Gamma$. Let us indicate by ${\cal B}_I$ and ${\cal B}_F$ the Borel $\sigma$-algebras of $I$ and $F$ respectively, and by ${\cal A}_I\equiv h_I^{-1}({\cal B}_I)$ and by ${\cal A}_F\equiv h_F^{-1}({\cal B}_F)$ the $\sigma$-algebras induced on $\Gamma$ by the two parametrizations. The following theorem, called factorization theorem, holds \cite{stocastic}: if a measurable map $f:I\rightarrow F$ exists such that $h_F=f\cdot h_I$, then ${\cal A}_F\subseteq {\cal A}_I$.

\vspace{5mm}
Let us proceed now to examine the measures. A parametrization $h:\Gamma\rightarrow R^n$ allows to induce on $\Gamma$ a measure $\mu$ on the basis of a measure $\nu$ defined on $R^n$, assuming $\mu(\Delta)\equiv \nu(h(\Delta))$, $\Delta\in h^{-1}({\cal B}^n)$. If $f:I\rightarrow F$ is a measurable map, it is possible to transfer a measure $\nu_I$ defined on $I$ into a measure $\nu_F$ defined on $F$ by assuming $\nu_F(\Delta)\equiv \nu_I(f^{-1}(\Delta))$, $\Delta\in {\cal B}_F$. According to the Radon-Nikodim theorem, every measure $\nu$ on $R^n$ that is absolutely continuous with respect to Lebesgue's measure can be represented in the form of a Lebesgue's integral of a measurable non-negative function $\rho$, called density:
\begin{equation}
\nu(\Delta)=\int\limits_{\Delta}^{}{\rho(x)dx},\; \Delta\in {\cal B}^n.
\label{3}
\end{equation}
If a measure $\nu_F$ on $F$ is induced by a measure $\nu_I$ on $I$ by means of a measurable map $f$, then the following relation holds between the corresponding densities $\rho_I$ and $\rho_F$:
\begin{equation}
\int\limits_{\Delta}\rho_{F}(y){\rm d}y=\int\limits_{f^{-1}(\Delta)}\rho_I(x){\rm d}x,\;\; \Delta\in{\cal B}_F.
\label{4}
\end{equation}
If the map $f$ is a local diffeomorphism, the following equation follows from relation (4):
\begin{equation}
	\rho_F(y)=\sum\limits_{i}\rho_I(x_{i})\left|\frac{\partial f_{i}^{-1}}{\partial y}\right|,
\label{5}
\end{equation}
where $\{x_i\}$ is the set of the $x\in I$ such that $f(x_i)=y$, $f_i$ is the restriction of $f$ to a neighborhood of $x_i$ that makes $f_i$ a diffeomorphism, and $\left|\frac{\partial f_i^{-1}}{\partial y}\right|$ is the jacobian of the transformation $f_i^{-1}$.

\vspace{.5cm}
Consider now the sets of paths in a hamiltonian system. For a more rigorous definition of hamiltonian system, including also the request that transformations be measurable, it is convenient to start from the definition of {\it abstract dynamical system} \cite{arnold}. An abstract dynamical system is a space $S$, called phase space, which has a $\sigma$-algebra ${\cal B}$, a class of measurable transformations $\{G_t\}_{t\in R}$ such that $G_0=1$ and $G_t\cdot G_u=G_{t+u}$, and an invariant measure $\nu$, i.e. a measure such that $\nu(G_t^{-1}(\Delta))=\nu(\Delta) \;\; \forall\Delta\in {\cal B}$, $\forall t\in R$. From the properties of class $\{G_t\}_{t\in R}$ one can easily infer that $G_t$ is invertible and that $G_t^{-1}=G_{-t}$. A path of the dynamical system is a map $\gamma:R\rightarrow S$ such that $\gamma(t)=G_t(s_0)$ for a given $s_0\in S$. 

An $N$-particle hamiltonian system is a dynamical system in which the phase space is $R^{6N}$, the $\sigma$-algebra is Borel's $\sigma$-algebra on $R^{6N}$, the class of trasformations is the one generated by Hamilton's equations and the measure is Lebesgue's measure, which Liouville's theorem guarantees to be invariant.

Consider the set of paths $\Gamma_T$ representing our simplified model of the universe. The two parametrizations $h_I:\Gamma_T\rightarrow I=R^{3N}$ and $h_F:\Gamma_T\rightarrow F=R^{3N}$, defined by setting $h_I(\gamma)\equiv \gamma_p(0)$ and $h_F(\gamma)\equiv \gamma_x(T)$, obviously correspond to the two kinds of boundary conditions $I$ and F. The map $h_I$ is bijective and therefore the atoms of ${\cal A}_I$ are the individual paths. Furthermore, assume without proof that the map $f\equiv h_F\cdot h_I^{-1}$, which associates with the initial momentum of a path its final position, is measurable; the factorization theorem leads to ${\cal A}_F\subseteq {\cal A}_I$. However, it is not true in general that ${\cal A}_I\subseteq {\cal A}_F$; in fact, since several paths can correspond to the same final position, $h_F$ is not bijective, and therefore the atoms of ${\cal A}_F$ are not the individual paths.

The term {\it classical measure}, and the symbol $\nu_{IC}$, indicate Lebesgue's measure on the set $I$ of initial momenta. In order to rigorously define the quantum measure $\nu_{FQ}$ on the set of final positions, we will proceed as follows: consider the hermitian operator $Q(T)\equiv e^{iHT}Qe^{-iHT}$; according to the spectral theorem, a spectral measure $E_T(\Delta)$, where $\Delta\in {\cal B}_F$, corresponds to this operator; let us define the measure $\nu_{FQ}$ by setting:
\begin{equation}
\nu_{FQ}(\Delta)\equiv \langle 0|E_T(\Delta)|0\rangle \;\;\Delta\in{\cal B}_F.
\label{misq1}
\end{equation}

Take $\mu_C$ to indicate the measure on $\Gamma_T$ induced by the parametrization $h_I$ starting from the measure $\nu_{IC}$ and $\mu_Q$ to indicate the measure induced by the parametrization $h_F$ starting from the measure $\nu_{FQ}$. Clearly, $\mu_C$ is defined on the $\sigma$-algebra ${\cal A}_I$ and $\mu_Q$ is defined on the $\sigma$-algebra ${\cal A}_F$. There are no theorems that guarantee, in general, the possibility of extending $\mu_Q$ to ${\cal A}_I$, i.e. of defining a measure $\mu'_Q$ on ${\cal A}_I$ such that $\mu'_Q(\Delta)=\mu_Q(\Delta) \; \forall \Delta\in {\cal A}_F$. Hereafter, however, we will assume that such an extension, albeit not univocal, does exist. This assumption is not fundamental to the development of the theory, but it will allow a clearer illustration of the relation between classical measure and quantum measure.

The parametrization $h_I$ can be considered as the element $h_0$ of the class of parametrizations $\{h_t\}_{t\geq0}$ from $\Gamma_T$ to $R^{6N}$ defined as follows: $h_t(\gamma)\equiv \gamma(t)$. One can easily see that the relation $h_{t+u}=G_u\cdot h_t$ holds, and that therefore $h^{-1}_t(\Delta)=h^{-1}_{t+u}(G_u(\Delta))$, $\Delta\subseteq R^{6N}$. Since $G_u$ is measurable, invertible and since its inverse is also measurable, we have $G_t({\cal B}^{6N})={\cal B}^{6N} \; \forall t$; therefore $h^{-1}_t({\cal B}^{6N})=h^{-1}_{t+u}({\cal B}^{6N})$ $\forall t,u\geq0$. Furthermore, since $h_0(\gamma)=(0_x,h_I(\gamma))$, it can be shown that $h^{-1}_t({\cal B}^{6N})={\cal A}_I$ $\forall t\geq0$.

A measure $\mu$ defined on ${\cal A}_I$, by means of the class of parametrizations $\{h_t\}_{\geq0}$ induces on $R^{6N}$ the class of measures $\{\nu_t\}_{t\geq0}$ defined as follows: $\nu_t(\Delta)\equiv \mu(h_t^{-1}(\Delta))$, $\Delta\in {\cal B}^{6N}$. From this definition it follows that $\nu_t(\Delta)=\nu_{t+u}(G_u(\Delta))$. If the class of measures $\{\nu_t\}_{t\geq0}$ is rapresented by means of a density $\eta (s,t)$ with respect to Lebesgue's measure on $R^{6N}$, a consequence of the relation $\nu_t(\Delta)=\nu_{t+u}(G_u(\Delta))$ and of the invariance of Lebesgue's measure for the transformations $\{G_t\}_{t\in R}$ is that the density $\eta$ satisfies the relation $\eta(s,t)=\eta(G_u(s),t+u)$; from the latter equation one finds that the density $\eta$ satisfies Liouville's equation. We shall use $\eta_C$ and $\eta'_Q$, respectively, to indicate the densities derived from the measures $\mu_C$ and $\mu'_Q$.

\vspace{5mm}
Let us summarize the results obtained in this section: the two types of boundary conditions $I$ and $F$ correspond to the two parametrizations $h_I(\gamma)\equiv \gamma_p(0)$ and $h_F(\gamma)\equiv \gamma_x(T)$ of $\Gamma_T$; these parametrizations induce on $\Gamma_T$ the two $\sigma$-algebras ${\cal A}_I$ and ${\cal A}_F$, with ${\cal A}_F\subset {\cal A}_I$; we have used $\nu_{IC}$ to indicate classical measure, i.e. the Lebesgue measure on the set of initial momenta I, and $\nu_{FQ}$ to indicate the quantum measure on the set of final positions $F$, which is defined by equation (\ref{misq1}). The measure $\nu_{IC}$ induce on ${\cal A}_I$ the measure $\mu_C$, whilst the quantum measure $\nu_{FQ}$ induce on ${\cal A}_F$ the measure $\mu_C$, which we extend to a measure $\mu'_Q$ on ${\cal A}_I$; the measures $\mu_C$ and $\mu'_Q$ on ${\cal A}_I$ induce on phase space two densities $\eta_C(s,t)$ and $\eta'_Q(s,t)$ which satisfy Liouville's equation.

\section{Comparison between classical measure and quantum measure.}

In the last section we defined the classical measure $\mu_C$ by deriving it from Lebesgue's measure $\nu_{IC}$ on the initial momenta. Since according to this definition the initial momenta of the particles are independent and mutually uncorrelated, the classical measure predicts initial conditions for the universe corresponding to the so-called ``molecular chaos". The measure $\nu_{IC}$ can be transferred by means of the map $f\equiv h_F\cdot h_I^{-1}$ onto a measure $\nu_{FC}$ defined on final position space by setting $\nu_{FC}(\Delta)\equiv \nu_{IC}(f^{-1}(\Delta))$, $\Delta\in {\cal B}_F$. Let us compute the expression for the density $\rho_{FC}$ that corresponds to the measure $\nu_{FC}$; if $f_i^{-1}(x)$ is the initial momentum of a path whose final position is $x$, where the index $i$ enumerates paths having the same final position, we have:
$$
f_i^{-1}(x)=\left.-\frac{\partial W_i(t_1,x_1,t_2,x_2)}{\partial x_1}\right|_{\begin{array}{ll} t_1=0,&x_1=0\\t_2=T,&x_2=x\end{array}},
$$
where $W_i(t_1,x_1,t_2,x_2)$ is the action between the points $(t_1,x_1)$ and $(t_2,x_2)$ computed along the $i$-th path {\cite{semiclassical}. Furthermore, since $\rho_{IC}=1$, we obtain from equation (5):
\begin{equation}
\rho_{FC}(x)=\sum_i\left|\frac{\partial W_i}{\partial x_1\partial x_2}\right|_{\begin{array}{ll} t_1=0,&x_1=0\\t_2=T,&x_2=x\end{array}}.
\label{7}
\end{equation}
The quantum measure $\nu_{FQ}$ defined on the final positions cannot instead be transferred directly on a measure $\nu_{IQ}$ defined on the $\sigma$-algebra ${\cal B}_I$ of the initial momenta because the map $f$ is not bijective. One can still define a measure $\nu'_{IQ}$ starting from an extension $\mu'_Q$ to the $\sigma$-algebra ${\cal A}_I$ of the measure $\mu_Q$, by setting $\nu'_{IQ}(\Delta)\equiv \mu'_Q(h_I^{-1}(\Delta))$, $\Delta\in {\cal B}_I$. Clearly, the measure $\nu'_{IQ}$ differs from Lebesgue's measure $\nu_{IC}$, and therefore it does not describe initial conditions of complete molecular chaos; rather, there will be correlations among the initial momenta of the particles. We will come back to this issue when dealing with preinteractive correlations.

By using the semi-classical approximation of the Feynman propagator $K(t_1,x_1,t_2,x_2)$, it is possible to express the quantum density $\rho_{FQ}$ on the final positions in a form which allows to compare it with the corresponding classical density $\rho_{FC}$. In semi-classical approximation, the propagator $\langle x|e^{-iHT}|0\rangle =K(0,0,T,x)$ becomes \cite{semiclassical}:
\begin{equation}
\langle x|e^{-iHT}|0\rangle =\frac{1}{(2\pi i)^{N/2}}=\left.\sum_i\left|\frac{\partial W_i}{\partial x_1\partial x_2}\right| ^{1/2} exp\left\{i\left(W_i-\frac{M_i\pi}{2}\right)\right\}\right|_{\begin{array}{ll} t_1=0,&x_1=0\\t_2=T,&x_2=x\end{array}},
\label{8}
\end{equation}
where $M_i$ is a phase factor. From equations (\ref{misq1}, \ref{7}, \ref{8}) one obtains:
\begin{equation}
\rho_{FQ}=\frac{1}{(2\pi)^N}(\rho_{FC}+\rho_{FI}),
\label{9}
\end{equation}
where $\rho_{FI}(x)$ in an interference term given by:
\begin{eqnarray}
\rho_{FI}(x)& = &\sum_{i\neq j} \left|\frac{\partial W_i}{\partial x_1\partial x_2}\right|^{1/2}\left|\frac{\partial W_j}{\partial x_1\partial x_2}\right|^{1/2}\times \nonumber \\
&& \times \left.exp\left\{i\left((W_i-W_j)-\frac{(M_i-M_j)\pi}{2}\right)\right\}\right|_{\begin{array}{ll} t_1=0,&x_1=0\\t_2=T,&x_2=x\end{array}}.
\label{10}
\end{eqnarray}
Thus, apart from the irrelevant proportionality factor $1/(2\pi)^N$, the difference between the two measures is given by the interference term $\rho_{FI}$.

\section{Limit for T$\rightarrow \infty$.}

The theory developed so far requires the existence of a final time $T$ for the universe. In this section we will take the limit for $T\rightarrow \infty$.

The path space $\Gamma_T$ is trivially extended to $\Gamma_{\infty}$, which we will indicate simply with $\Gamma$. As to the parametrization $h_F$, it is not possible to simply take the limit for $T\rightarrow \infty$ of $\gamma_x(T)$, since such a limit in general does not exist. However, the limit of $\gamma_x(T)/T$ for $T\rightarrow \infty$ does exist, as shown later. For reasons which will become evident, instead of the expression $\gamma_x(T)/T$ we will consider the expression $m\gamma_x(T)/T$, where by $m\gamma_x(T)$ we mean $(m_1{\bf \gamma}_{1x}(T),...,m_N{\bf\gamma}_{Nx}(T))$, where ${\bf \gamma}_{ix}(T)$ is the position of the $i$-th particle at the time $T$. We therefore define the notion of {\it p-limit} of a path $\gamma\in \Gamma$, indicated by the symbol $p-\lim\gamma$, in the following way:
\begin{equation}
p-\lim\gamma \equiv \lim_{t\rightarrow + \infty}\frac{m\gamma_x(t)}{t}
\label{11}
\end{equation}
It is easy to prove the two following lemmas:
\begin{itemize}
\item if $\gamma_p(t)\rightarrow p$ for $t \rightarrow +\infty$ then $p-\lim\gamma=p$;
\item if a vector $v\in R^{3N}$ exists such that $\|\gamma_x(t)-vt\|$ is bounded for $t\geq0$, then $p-\lim\gamma=mv$.
\end{itemize}
The first lemma states that if the momentum of the path $\gamma$ has a limit for $t\rightarrow +\infty$, then that limit corresponds to the p-limit; hence the name ``p-limit" given to the limit (\ref{11}). The second lemma states that if the distance between the path position and a point moving with constant velocity $v$ is bounded for every $t\geq0$, then the p-limit exists and is $mv$ \footnote{In particular, if $v=0$, i.e. for the bound states of the system, the p-limit is 0.}. This means that the p-limit can exist even if the limit of the momentum for $t\rightarrow +\infty$ does not exist, provided that the required condition holds. In order to use the p-limit as a parametrization, however, it must exist for all the paths of the space $\Gamma$. We will not prove this statement in general, but we will show a result of the classical scattering theory which makes this conjecture plausible \cite{scattering}.

Let $\alpha=\{F_1,...,F_n\}$ be a partition of the $N$ particles in $n$ subsets $F_1,...,F_n$. In scattering theory, a partition of this kind is known as a {\it channel}, and the subsets $F_k$ are called {\it fragments} of the channel. For every fragment $F_k$, we will indicate by ${\bf X}_k$ and ${\bf V}_k$ the position and velocity, respectively, of the center of mass of the system consisting of the particles that belong to the fragment $F_k$. Furthermore, we will indicate by $S_{\alpha+}\subseteq S$ the subset of phase space crossed by those paths for which the following properties hold:
\begin{itemize}
\item the velocities ${\bf V}_k(t)$ of each fragment have limit ${\bf V}_k(\infty)$ for $t\rightarrow +\infty$, with ${\bf V}_k(\infty)\neq{\bf V}_j(\infty)$ if $k\neq j$;
\item every fragment $F_k$ is a bound system, i.e. $\|{\bf X}_k(t)-{\bf \gamma}_{ix}(t)\|$ is bounded for $t\geq0$ and $i\in F_k$.
\end{itemize}
Those properties mean that the paths of $S_{\alpha+}$ describe for $t\rightarrow +\infty$ the $N$ particles that break up into $n$ fragments identified by the partition $\alpha$ which are bound and move away from each other, keeping a constant momentum. It is easy to see that all the paths defined by $S_{\alpha+}$ admit a p-limit, and that the p-limit of the $i$-th particle belonging to the fragment $F_k$ is $m_i{\bf V}_k(\infty)$. If the potentials $V_{ij}$ have a finite range, one can prove that Lebesgue's measure of the set of phase space points that do not belong to an $S_{\alpha+}$ for some channel $\alpha$ is null, i.e. $\nu_L(S\setminus \cup_{\alpha}S_{\alpha+})=0$.

On the basis of this result we will use the p-limit as a parameter, by setting $h_F(\gamma)\equiv p-\lim\gamma$.

\vspace{.5cm}
In order to take the limit of the quantum measure $\nu_{FQ}$, we introduce the operators $P_+$ and $P_-$:
\begin{equation}
P_{\pm}\equiv\lim_{t\rightarrow \pm\infty}\frac{mQ(t)}{t},
\label{12}
\end{equation}
where once again $mQ$ indicates $(m_1{\bf Q}_1,...,m_N{\bf Q}_N)$. If the limit exists, the operators $P_{\pm}$ are hermitian, since the operators $Q(t)$ are hermitian for every $t$. Let us study the limit (\ref{12}) in the case of a particle of mass $m$ in a central potential $V(\|{\bf x}\|)$. The Hamiltonian will be:
$$
H=-\frac{\Delta^2}{2m}+V(\|{\bf x}\|)=H_0+V.
$$
We initially compute the limit (\ref{12}) in the case of lack of potential. Using the Feynman propagator for the free particle:
\begin{equation}
K({\bf x},{\bf y},t)=\left(\frac{m}{2\pi it}\right)^{3/2}\exp\frac{im({\bf x}-{\bf y})^2}{2t},
\label{14}
\end{equation}
we have:
\begin{eqnarray}
&&\left(\frac{me^{iH_0t}{\bf Q}e^{-iH_0t}}{t}\psi\right)({\bf x})= \left(\frac{m}{2\pi t}\right)^3
\int \exp\frac{-im({\bf x-y})^2}{2t}\frac{m{\bf y}}{t}\exp\frac{im({\bf y-z})^2}{2t}\psi({\bf z})d^3yd^3z \nonumber \\
&&= \left(\frac{1}{2\pi}\right)^3 \exp\frac{-im{\bf x}^2}{2t} \int {\bf p}\exp[-i{\bf p}({\bf z-x})]\exp\frac{im{\bf z}^2}{2t}\psi({\bf z})d^3pd^3z \nonumber \\
&&= \exp\frac{-im{\bf x}^2}{2t}(-i)\nabla_x\left(\exp\frac{im{\bf x}^2}{2t}\psi({\bf x})\right)=
\frac{m{\bf x}}{t}\psi({\bf x})-i\nabla_x\psi({\bf x}) \nonumber \\
&&= \left(\frac{m{\bf Q}}{t}\psi\right)({\bf x})+
({\bf P}\psi)({\bf x}),\nonumber
\end{eqnarray}
from which
\begin{equation}
\lim_{t\rightarrow\pm\infty}\frac{me^{iH_0t}{\bf Q}e^{-iH_0t}}{t}={\bf P}.
\label{16}
\end{equation}
It is easy to compute the limit (\ref{12}) even in the presence of a potential which allows to define the M\"{o}ller operators $\Omega_{\pm}$:
$$
\Omega_{\pm}=\lim_{t\rightarrow\pm\infty} e^{iHt}e^{-iH_0t}.
$$
We use $E_B$ and $E_S$ to indicate, respectively, the projectors on the space of bounded states and on the space of scattering states. Due to the property of asymptotic completeness, the relation $E_B+E_S=I$ holds. Consider initially the scattering states. We have:
$$
\lim_{t\rightarrow\pm\infty}\frac{m{\bf Q}(t)}{t}E_S= \lim_{t\rightarrow\pm\infty}e^{iHt} e^{-iH_0t} e^{iH_0t}\frac{m{\bf Q}}{t} e^{-iH_0t} e^{iH_0t} e^{-iHt}E_S=\Omega_\pm{\bf P}\Omega_\pm^\dag E_S.
$$
As regards the bounded states, consider a bounded state $\psi_n$ of the following type:
$$
\psi_n=\sum\limits_{i=1}^n a_i\phi_i,
$$
where the states $\phi_i$ are eigenstates of the Hamiltonian with discrete eigenvalue $E_i$. We have:
$$
\|e^{iHt}\frac{m{\bf Q}}{t}e^{-iHt}\psi_n\|=\frac{m}{t}\|\sum\limits_{i=1}^n a_i\exp(-iE_i t){\bf Q}\phi_i\| \leq\frac{m}{t}\sum\limits_{i=1}^n \| a_i{\bf Q}\phi_i\|,
$$
which vanishes for $t\rightarrow \pm \infty$, from which ${\bf P}_{\pm}\psi_n=0$. For a generic bounded vector $\psi$ there is a sequence of vectors $\psi_n$ such that $\psi_n\rightarrow \psi$ for $n\rightarrow \infty$; furthermore, obviously ${\bf P}_\pm \psi_n\rightarrow 0$ for $n\rightarrow \infty$. From the two limits $\psi_n\rightarrow \psi$ and ${\bf P}_\pm\psi_n\rightarrow 0$ one infers ${\bf P}_\pm\psi=0$ only if the operator ${\bf P}_\pm$ is closed \cite{jauch}, which we assume to be true without proving it. Hence ${\bf P}_\pm E_B=0$.

Finally, we have:
\begin{equation}
{\bf P}_\pm=\Omega_\pm{\bf P}\Omega_\pm^\dag E_S.
\label{21}
\end{equation}
The operators ${\bf P}_\pm$ commute with the Hamiltonian. In fact we have:
$$
H{\bf P}_\pm=H\Omega_\pm{\bf P}\Omega_\pm^\dag E_S=
\Omega_\pm H_0{\bf P}\Omega_\pm^\dag E_S=
\Omega_\pm{\bf P}H_0\Omega_\pm^\dag E_S=
\Omega_\pm{\bf P}\Omega_\pm^\dag HE_S={\bf P}_\pm H.
$$
We will not extend the study of the limit (\ref{12}) to the $N$-particle case, and we assume that in that case also the operators $P_\pm$ exist and commute with the Hamiltonian.
In order to define the measure $\nu_{FQ}$ we use the spectral measure associated with the operators $P_+$, which we indicated with $E_+(\Delta)$. Let then:
\begin{equation}
\nu_{FQ}(\Delta)\equiv\langle0| E_+(\Delta)|0\rangle, \; \Delta\in {\cal B}_F.
\label{23}
\end{equation}
One can verify the formal consistency of equation (\ref{23}) by computing the equation in the case of vanishing potential, where $P_+=P$. We have:
$$
\langle0| E_+(\Delta)|0\rangle=\int \langle0|p_1\rangle dp_1
\langle p_1|E_+(\Delta)|p_2\rangle dp_2\langle p_2|0\rangle=
\int \chi_\Delta (p_1)\delta(p_2-p_1)dp_1dp_2=\nu_{FL}(\Delta),
$$
where $\chi_\Delta$ is the indicator function of the set $\Delta$, and $\nu_{FL}$ is the Lebesgue measure on $F=R^{3N}$.

\vspace{5mm}
To summarize, the boundary conditions of the universe on which we define the quantum measure are the p-limit of the paths, defined by equation (\ref{11}), and the quantum measure is given by relation (\ref{23}), where $E_+(\Delta)$ is the spectral measure associated with the operator $P_+$ defined by equation (\ref{12}).

\section{The two-slit experiment.}

Let us go back to examining the physical aspects of the quantum-classical theory by studying the two-slit experiment in the light of this theory. Consider the experimental apparatus illustrated in figure 1, which is used to produce electron diffraction:

\begin{center}
\unitlength=1mm
\begin{picture}(120,35)
\put(54,5){\line(1,0){10}}
\put(54,30){\line(1,0){10}}
\put(58,17){\circle*{1}}
\put(112,5){\line(0,1){23}}
\put(0,16){\line(1,0){7}}
\put(7,16){\line(0,1){1}}
\put(7,18){\line(0,1){1}}
\put(0,19){\line(1,0){7}}
\put(0,16){\line(0,1){3}}

\put(47,2){\makebox(6,6){$E_2$}}
\put(47,27){\makebox(6,6){$E_1$}}
\put(50,13){\makebox(6,6){$F$}}
\put(105,1){\makebox(6,6){$H$}}

\put(0,10){\makebox(6,6){$S$}}

\end{picture}

Fig. 1
\end{center}

Here S is the electron source, $F$ is a thin conducting filament which crosses at right angles the plane of the figure and is set to a positive potential with respect to the two electrodes $E_1$ and $E_2$, and H is a screen consisting of a photographic plate. Due to the electrostatic field generated by the filament, the electrons emitted by the source are deflected and give rise to interference fringes on the screen. Let us see how this phenomenon can be explained in the framework of the quantum-classical theory.

According to this theory, the electrons travel between the source and the screen along classical paths whose statistical distribution derives from the measure on the boundary conditions of the universe. This measure is such that it does not determine a uniform distribution of the momenta of the electrons when they are emitted by the source; rather, it determines a distribution of the electron impact points on the screen which corresponds to the interference fringes. To make this argument more precise and consistent, however, we need to make two conjectures.

Consider the four paths shown in figure 2:

\begin{center}
\unitlength=1mm

\begin{picture}(120,21)
\put(58,12){\circle*{1}}
\put(112,0){\line(0,1){23}}

\put(112,12){\circle*{1}}
\put(112,9){\makebox(6,6){$P$}}

\put(112,18){\circle*{1}}
\put(112,15){\makebox(6,6){$P'$}}
\put(20,13){\makebox(6,6){$\gamma_1$}}

\bezier{600}(7,12)(58,16)(112,12)
\put(0,11){\framebox(7,3){ }}
\end{picture}

\vspace{5mm}
\unitlength=1mm
\begin{picture}(120,21)
\put(58,12){\circle*{1}}
\put(112,0){\line(0,1){23}}

\put(112,12){\circle*{1}}
\put(112,9){\makebox(6,6){$P$}}

\put(112,18){\circle*{1}}
\put(112,15){\makebox(6,6){$P'$}}
\put(20,13){\makebox(6,6){$\gamma_2$}}

\bezier{600}(7,12)(58,8)(112,12)
\put(0,11){\framebox(7,3){ }}
\end{picture}

\vspace{5mm}
\unitlength=1mm
\begin{picture}(120,21)
\put(58,12){\circle*{1}}
\put(112,0){\line(0,1){23}}

\put(112,12){\circle*{1}}
\put(112,9){\makebox(6,6){$P$}}

\put(112,18){\circle*{1}}
\put(112,15){\makebox(6,6){$P'$}}
\put(20,13){\makebox(6,6){$\gamma_3$}}

\bezier{600}(7,12)(58,16)(112,18)
\put(0,11){\framebox(7,3){ }}
\end{picture}

\vspace{5mm}
\unitlength=1mm
\begin{picture}(120,21)
\put(58,12){\circle*{1}}
\put(112,0){\line(0,1){23}}

\put(112,12){\circle*{1}}
\put(112,9){\makebox(6,6){$P$}}

\put(112,18){\circle*{1}}
\put(112,15){\makebox(6,6){$P'$}}
\put(20,13){\makebox(6,6){$\gamma_4$}}

\bezier{600}(7,12)(58,6)(112,18)
\put(0,11){\framebox(7,3){ }}
\end{picture}

Fig. 2
\end{center}

where P and P' are two distinct points on the screen which correspond respectively to a maximum and a minimum of the interference fringes; these four paths must be considered not simply as electron paths, but rather as overall paths of the universe. The conjectures we make are the following:
\begin{itemize}{}%
\item[$C_1)$] paths $\gamma_1$ and $\gamma_2$ have a p-limit which is different from the p-limit of paths $\gamma_3$ and $\gamma_4$.
\item[$C_2)$] paths $\gamma_1$ e $\gamma_2$ have the same p-limit; so do paths $\gamma_3$ and $\gamma_4$.
\end{itemize}
Let us give physical reasons for the two conjectures. As regards conjecture $C_1$, we observe that in order for two paths of the universe to have different p-limits it is sufficient for a single particle in the universe to have a different p-limit in the two paths. For instance, if in the two paths a photon is emitted in space in two different directions, the two paths have a different p-limit. It is therefore reasonable to think that if two paths are able to mutually differ macroscopically, they determine in the surrounding environment modifications that render their differences permanent and therefore lead them to have different p-limits. This holds, for instance, for paths $\gamma_1$ and $\gamma_3$, in which the electron produces, in both cases, the permanent blackening of two macroscopically distinct silver grains. In other words, conjecture $C_1$ states that the measurement of which point of the screen was struck by the electron can be made, at least from the theoretical point of view, at the instant $T\approx\infty$ by measuring the quantity $m\gamma_x(T)/T$, i.e. by means of a position measurement at the final instant of the universe. Conjecture $C_1$ can therefore be considered the classical version of conjecture $Q_1$ formulated earlier. We can express it in a more general way in the following form:
\begin{quote}
{\it If two paths of the universe attain macroscopic mutual differences, they have different p-limits; equivalently, if two paths have the same p-limit, they are macroscopically indistinguishable.}
\end{quote}
Earlier we observed that the definition of the final position of a path (and consequently of its p-limit) do not univocally identify the path. If conjecture $C_1$ is true, this fact is not a physically relevant problem, since such paths, for all physical purposes, will be indistinguishable.

As regards conjecture $C_2$, consider the two pairs of paths $\{\gamma_1,\gamma_2\}$ and $\{\gamma_3,\gamma_4\}$. The two paths of each pair never reach the point of differing macroscopically from each other and it is therefore reasonable to think they have the same p-limit. Incidentally, we would like to point out that this conjecture cannot be inferred from conjecture $C_1$, since the latter {\it does not state} that if two paths are macroscopically indistinguishable they have the same p-limit. Due to conjecture $C_2$, the two paths of each pair mutually interfere in determining the measure, as shown hereafter; furthermore, individual paths are not measurable with respect to the $\sigma$-algebra ${\cal A}_F$: only the sets $\{\gamma_1,\gamma_2\}$ and $\{\gamma_3,\gamma_4\}$ are measurable.

In view of these conjectures, and letting $p\equiv p-\lim\gamma_1=p-\lim\gamma_2$ and $p'\equiv p-\lim\gamma_3=p-\lim\gamma_4$, the interference fringes appear if the densities $\rho_{FQ}(p)$ and $\rho_{FQ}(p')$ are different and proportional to the fringe intensities. Due to equation (\ref{9}), which gives the semiclassical approximation of the quantum measure $\rho_{FQ}$, one can state that the two paths of each pair mutually ``interfere" in determining the measure $\rho_{FQ}$. One should bear in mind that this interference acts only in determining the measure, whilst the electron travels physically along one of the two paths.

It is furthermore easy to understand the mechanism by which any attempt to observe which of the two paths the electron actually travels along disrupts the interference: such an observation, in fact, can only consist in amplifying to the macroscopic level the differences between the two paths, thus giving them different p-limits and consequently making them no longer interfere with each other.


\section{The scattering process.}

Another physical example which we analyze in the light of the quantum-classical theory is the scattering process\footnote{ Although this example does not use exactly the simplified model of the universe, completely fits the spirit of the quantum-classical theory.}. Consider a Hamiltonian system consisting of a particle of mass $m$ in a three-dimensional space. The phase space is $R^6$. The particle moves under the action of a central potential $V(\|{\bf x}\|)$ which vanishes quickly enough for $\|{\bf x}\|\rightarrow\infty$. The Hamiltonian is:
$$
H=\frac{{\bf p}^2}{2m}+ V(\|{\bf x}\|).
$$
The set of paths $\Gamma$ we consider is the set consisting of those paths which have a fixed asymptotic initial momentum ${\bf p}_I$, which we choose to be oriented in the positive sense of the $z$ axis: 
$$
\Gamma=\{\gamma:\lim_{t\rightarrow-\infty}\gamma_p(t)=p_I{\bf \hat{u}}_z\}.
$$
Each path identifies three variables: the impact parameter $b$, the scattering angle $\theta$ and the angle $\phi$ between the $x$ axis and the plane of the orbit. Let us define the parametrizations $h_I$ and $h_F$ by letting $h_I(\gamma)=(b,\phi)$ and $h_F(\gamma)=(\theta,\phi)$. The function $f:(b,\phi) \mapsto (\theta,\phi)$ is well-defined, and we assume it to be measurable in this case also. Therefore we have $h_F=f\cdot h_I$, from which it follows, due to the factorization theorem, that ${\cal A}_F\subseteq {\cal A}_I$. The function $f$ is not invertible, since there can be orbits with different impact parameters but with the same scattering angle; this can happen, for instance, with a suitable attractive potential, in which different orbits with the same scattering angle correspond to a different number of revolutions performed by the particle around the center of attraction. We will therefore have ${\cal A}_F\subset{\cal A}_I$.

Consider two measures $\nu_I$ and $\nu_F$ defined respectively on the ranges $I$ and $F$ of $h_I$ and $h_F$, and express them by means of the two densities $\rho_I(b,\phi)$ and $\rho_F(\theta,\phi)$. We have:
\begin{equation}
\nu_I(\Delta_I)=\int_{\Delta_I}\rho_I(b,\phi)bdbd\phi, \; \Delta_I\in {\cal B}_I;
\label{26}
\end{equation}
$$
\nu_F(\Delta_F)=\int_{\Delta_F}\rho_F(\theta,\phi)\sin\theta d\theta d\phi, \;
\Delta_F\in {\cal B}_F.
$$
Here $\rho_I$ corresponds to the statistical distribution of the paths on the incident plane, and $\rho_F$ to their statistical distribution in the solid angle $d\Omega=\sin\theta d\theta d\phi$.

If $\nu_F$ is induced by $\nu_I$, from equation (\ref{5}) it follows that:
\begin{equation}
\rho_F(\theta,\phi)=\sum_i\rho_I(b_i,\phi)\frac{b_i}{\sin\theta}\left|\frac{\partial b_i}{\partial\theta}\right|,
\label{27}
\end{equation}
where $\{b_i\}$ is the set of values of the impact parameter which determine the same scattering angle $\theta$. The two measures $\nu_I$ and $\nu_F$ induce on $\Gamma$ two measures $\mu_I$ and $\mu_F$ defined respectively on ${\cal A}_I$ and ${\cal A}_F$ in the manner described earlier.

\vspace{.5cm}
Consider now the relation existing between cross-section and measure on path space. The cross-section $\sigma(\theta,\phi)$ is operatively defined in the following way:
\begin{equation}
\sigma(\theta,\phi)d\Omega=\frac{{\rm number\>of\> particles\> scattered\> for\> unit\> time\> in\> the\> solid\> angle\> d\Omega}}{{\rm number\> of\> incoming\> particles\> for\> unit\> time\> and\> unit\> area}}.
\label{28}
\end{equation}
Suppose that a measure $\mu$ is defined on the space $\Gamma$ without specifying - for the time being - on which of the two $\sigma$-algebras ${\cal A}_I$ and ${\cal A}_F$ it is actually defined. In order to express the cross-section in terms of this measure, we set the following natural correspondence: {\it the number of particles per unit time traveling along a given set of paths is proportional to the measure of that set}. Equation (\ref{28}) can then be expressed in the following way:
\begin{equation}
\sigma(\theta,\phi)d\Omega=\frac{{\rm measure\> of\> the\> paths\> scattered\> in\> the\> solid\> angle\> d\Omega}}{{\rm measure\> of\> incoming\> paths\> for\> unit\> area}}.
\label{29}
\end{equation}
Let us examine the expression at the numerator. The set of paths scattered in the solid angle $d\Omega$ is $h_F^{-1}(\{d\theta d\phi\})\in {\cal A}_F$, and its measure is $\rho_F(\theta,\phi)\sin\theta d\theta d\phi$.
We can compute the expression at the denominator in the following way. If $\Delta\in {\cal A}_F$ is a set of paths and we use $\nu_{IC}$ to indicate Lebesgue's measure on the incidence plane, the quantity $\nu_{IC}(h_I(\Delta))$ is the area of the incident surface of the set of paths $\Delta$. Formally, the quantity at the denominator, which we indicate with $\bar\mu$, should then be:
$$
\bar\mu=\frac{\mu(\Gamma)}{\nu_{IC}(h_I(\Gamma))}.
$$
However, such a quantity is generally of the form $\infty/\infty$ and therefore it is not defined. Accordingly, it is necessary to define it by means of a limit. Consider for instance a sequence $\{\Delta_n\}$ of subsets of ${\cal A}_F$ such that $\Delta_n\subset \Delta_{n+1}$ and such that $\cup_n\Delta_n=\Gamma$. We can then define $\bar\mu$ in the following way:
\begin{equation}
\bar\mu=\lim_{n\rightarrow\infty}\frac{\mu(\Delta_n)}{\nu_{IC}(h_I(\Delta_n))}.
\end{equation}
We shall not deal here with the issue of the existence of this limit or with the issue of its dependence on the sequence $\{\Delta_n\}$. In conclusion, for the scattering cross-section we have:
\begin{equation}
\sigma(\theta,\phi)=\frac{\rho_F(\theta,\phi)}{\bar\mu}.
\label{33}
\end{equation}

From the above arguments and from equation (\ref{33}) one infers that: (i) in order to express the cross-section in terms of a measure on the path space it is sufficient for that measure to be defined on the $\sigma$-algebra ${\cal A}_F$, not on the more refined $\sigma$-algebra ${\cal A}_I$; (ii) the cross-section essentially corresponds to the density of the measure $\nu_F$.

\vspace{.5cm}
The hypothesis that is at the basis of the derivation of the classical cross-section $\sigma_C$ is that the incoming particles are uniformly distributed (with respect to Lebesgue's measure) on the incidence plane. This hypothesis can be expressed mathematically by assuming that the measure $\mu$ is induced by $\nu_{IC}$, i.e. by Lebesgue's measure on the incidence plane. Clearly, one can derive from this that $\rho_{IC}=1$, $\bar\mu=1$ and can derive the following from equation (\ref{27}):
\begin{equation}
\sigma_C(\theta,\phi)=\sum_i\frac{b_i}{\sin\theta}\left|\frac{\partial b_i}{\partial\theta}\right|,
\label{34}
\end{equation}
which is the well-known formula for classical scattering cross-section.

One can instead obtain the quantum cross-section $\sigma_Q$ by directly assuming:
\begin{equation}
\rho_{FQ}(\theta,\phi)=\sigma_Q(\theta,\phi).
\label{35}
\end{equation}
The quantum cross-section can therefore be obtained in a classical dynamical framework simply by defining appropriately the measure on path space.

As shown earlier, it is possible to associate with the measure $\nu_{FQ}$ defined by relation (\ref{35}) a measure $\nu'_{IQ}$ describing the statistical distribution of the particles on the incidence plane in the quantum case. Such a measure is clearly different from Lebesgue's measure $\nu_{IC}$, otherwise one would obtain the classical cross-section again. This means that one can obtain the quantum cross-section if one admits that the incoming particles are not uniformly distributed on the incidence plane, and that their distribution is such that it determines precisely the value (\ref{35}) for $\rho_{FQ}$. Since this value clearly depends on the form of the potential, the incoming particles must be distributed on the incidence plane in a way which is correlated with the form of the potential, and this must be true even long before interacting with the potential itself. This is an example of preinteractive correlation, a phenomenon which we will study more thoroughly in section 12.

\section{Some possible objections to the quantum-classical theory.}

In this section we will consider some possible objections to the fact that the quantum-classical theory can account for the phenomena predicted by orthodox quantum theory. For the sake of simplicity, in this discussion we assume that the universe has a final time $T$. We briefly summarize the main points of the quantum-classical theory and of quantum theory, the latter intended as a universal theory:
\begin{enumerate}
\item {\it Quantum-classical theory}: the evolution of the universe is represented by a path connecting the points $(0,0)$ and $(x,T)$ and satisfying the least action principle. The statistical distribution $\rho_{FQ}$ of the final position $x$ is defined by the equation $\rho_{FQ}(x)\equiv |\langle x|e^{-iHT}|0\rangle|^2$.
\item {\it Quantum mechanics}: the wave function of the universe at every time $t$ is $\psi(x,t)\equiv \langle x|e^{-iHt}|0\rangle$; it is possible to extract from the wave function a classical representation of the evolution by using, for example, ``Consistent histories approach to quantum mechanics" techniques.
\end{enumerate}

{\it First objection}. We have seen that quantum-classical theory allows to define, albeit not univocally, a probability density $\eta'_Q(x,p,t)$ which satisfies Liouville's equation. By integrating over momenta it is possible to obtain from $\eta'_Q(x,p,t)$ a space probability density $\rho'_Q(x,t)$:
$$
\rho'_Q(x,t)\equiv\int\eta'_Q(x,p,t)dp.
$$

If quantum-classical theory were compatible with quantum mechanics, the relation $\rho'_Q(x,t)=|\psi(x,t)|^2$ should hold for every $x$ and $t$. This is clearly impossible, because whilst the evolution of $\rho'_Q(x,t)$ is determined by Liouville's equation, the evolution of $|\psi(x,t)|^2$ is determined by Schr\"{o}dinger's equation, which is structurally different from Liouville's.

Answer: the equation $\rho'_Q(x,t)=|\psi(x,t)|^2$ should not be set as a condition for the compatibility of the two theories. Our perception of quantum phenomena is not direct but is always mediated by macroscopic devices: no one has ever directly measured or observed the wave function. In order to consider the two theories compatible, {\it it is therefore sufficient that $\rho'_Q(x,t)$ and $|\psi(x,t)|^2$ describe the same macroscopic configuration of measuring instruments, i.e. it is sufficient that they are equal only ``macroscopically"}. It seems appropriate, in this regard, to quote Stapp \cite{stapp} on the issue of quantum mechanics and wave function: ``How can a theory which is {\it fundamentally} a procedure by which gross macroscopic creatures, such as human beings, calculate predicted probabilities of what they will observe under macroscopically specified circumstances ever be claimed to be a complete description of physical reality?"

Discussing this objection allows us to highlight some analogies and differences between the quantum-classical theory and Bohm's causal theory \cite{bohm}. The latter, too, predicts that particles should follow definite paths and that the statistical distribution of these paths should give rise to a density $\rho(x,t)$. For this density, however, the equation $\rho(x,t)=|\psi(x,t)|^2$ must hold for every $x$ and $t$. In order for this to happen, the Hamiltonian that rules particle evolution must include an additional term, the so called quantum potential, which is attributed to the interaction between the particle and the wave function. If $\rho(x,t)=|\psi(x,t)|^2$ is not a requirement, the need for quantum potential disappears and we obtain the quantum-classical theory.

{\it Second objection}. It is well-known that the wave function is different from 0 also in regions not reached by classical paths, such as for instance in the vicinity of a caustic \cite{path}. What happens then if the final state of the universe is a point $x$ which cannot be reached classically?

Answer: this is certainly a significant objection to which, for the time being, we can only give indicative answers. First of all, it should be specified that this objection has nothing to do with the tunnel effect, which is a fixed-energy phenomenon which does not fit this context. For instance, the tunnel effect is not considered an impediment to the possibility of obtaining the Feynman propagator, calculated by the formalism of Feynman's path integrals, by summing only classical paths \cite{path}.

Perhaps when taking the limit for $T\rightarrow\infty$ the classically-reachable regions and the quantistically-reachable ones ultimately coincide; this would solve the objection. On the other hand, taking the limit may produce the opposite effect, causing regions that can be reached classically at a finite time to become classically unreachable. To understand the importance of this objection, it is also necessary to understand the percentage of final positions that cannot be reached classically with respect to the total. In any case, in order to answer this objection the asymptotic properties of the wave function must be investigated more carefully.

{\it Third objection}. How can the quantum-classical theory explain phenomena such as the photoelectric effect, the Compton effect, or atomic spectra?

Answer: we stated that the quantum-classical theory, in the formulation introduced in this paper, is based on the dynamical scheme of a hamiltonian system of particles, in which particles do not have an internal structure and in which the number of particles is conserved. For this reason, the quantum-classical theory deals exclusively with phenomena predicted by Schr\"{o}dinger's equation in its most elementary form. The above mentioned phenomena instead involve the interaction of electrons with photons and thus belong to a broader dynamical scheme. The extension of the quantum-classical theory to this class of phenomena will probably be the subject of a future paper. The terms of this extension - and the significant fact that it probably entails abandoning the deterministic character of the quantum-classical theory - are mentioned briefly in section 13.

{\it Fourth objection}. The EPR paradox forbids a local deterministic theory from reproducing the results predicted by quantum mechanics. The quantum-classical theory, in addition to being a deterministic theory, is also a local theory, since it predicts that the mutual interaction of particles becomes negligible when they are distant enough from each other.

We will discuss the EPR paradox in the next section.


\section{Preinteractive correlations and the EPR paradox.}

This section studies the phenomenon of preinteractive correlations and their possible role in overcoming the EPR paradox. The term {\it preinteractive correlations} was borrowed from Price, who has discussed the topics extensively \cite{price1}\cite{price2}\cite{price3}\cite{price4}; Prigogine speaks of precollisional interactions \cite{prigogine}.

The mechanism giving rise to preinteractive correlations is simple, and can be illustrated by studying a simple collision between two particles. Consider a one-dimensional universe, consisting of two point-like particles 1 and 2 of mass $m_1$ and $m_2$, with $m_1\gg m_2$. The particles interact with each other by means of an elastic collision, and in the approximation $m_1\gg m_2$ the velocity of particle 1 is not changed by the collision. Suppose that particle 1 is still and that particle 2 collides with it coming from the left. Accordingly, we have $p_1=0$ and $x_2\leq x_1$. Furthermore, before the collision we have $p_2=m_2v>0$, whilst after the collision we have $p_2=-m_2v$. If $(x_1,0,x_2,m_2v)$ is an element of the phase space that describes the particles before the collision, for $t>(x_1-x_2)/v$, i.e. after the collision, we have that $G_t(x_1,0,x_2,m_2v)=(x_1,0,x_2-vt+2x_1,-m_2v)$.

Section 6 showed that a measure on boundary conditions induces a density $\eta(s,t)$ on the phase space which satisfies Liouville's equation. Accordingly, let us develop this example by using phase space densities instead of measures. Consider therefore the following three densities, which represent collisions with the previously described features:
\begin{equation}
\begin{array}{ll}
\eta_+(x_1,p_1,x_2,p_2,t)= & \left\{
\begin{array}{ll}
\rho_1(x_1)\delta(p_1)\rho_2(x_2-vt)\delta(p_2-mv) & for\; t<t_I \\
\rho_1(x_1)\delta(p_1)\rho_2(-x_2-vt+2x_1)\delta(p_2+mv) & for\; t>t_F
\end{array}\right.\nonumber \\
%
\eta_-(x_1,p_1,x_2,p_2,t)= & \left\{
\begin{array}{ll}
\rho_1(x_1)\delta(p_1)\rho_2(x_2-vt-2x_1)\delta(p_2-mv) & for\; t<t_I \\
\rho_1(x_1)\delta(p_1)\rho_2(-x_2-vt)\delta(p_2+mv) & for\; t>t_F
\end{array}\right. \\
%
\eta_\pm(x_1,p_1,x_2,p_2,t)= & \left\{
\begin{array}{ll}
\rho_1(x_1)\delta(p_1)\rho_2(x_2-vt-x_1)\delta(p_2-mv) & for\; t<t_I \\
\rho_1(x_1)\delta(p_1)\rho_2(-x_2-vt+x_1)\delta(p_2+mv) & for\; t>t_F
\end{array}\right.\nonumber
\end{array}
\end{equation}
where $\rho_1$ and $\rho_2$ are two space distributions which vanish outside a bounded region which is centered around 0 and where $t_I<0$ and $t_F>0$ are two instants before and after the collision, at a sufficient distance from it. It is easy to see that the three densities correctly satisfy the relation $\eta(s,t)=\eta(G_u(s),t+u)$. All three densities describe the heavy particle as being spatially distributed according to $\rho_1$ in a bounded region which is centered around the origin. They differ, on the other hand, in the space distribution of the light particle. In the case of $\eta_+$, before the collision this distribution does not depend on the position of the heavy particle, whilst it depends on it after the collision; this case describes a normal postinteractive correlation and the lack of preinteractive correlations. In the second case, instead, the space distribution of the light particle depends on the position of the heavy particle {\it before} the collision and this dependence is removed by the collision; this case has only a preinteractive correlation. Finally, the third case has both a preinteractive correlation and a postinteractive correlation.

The above example should clarify the nature of preinteractive correlations, and particularly the fact that no dynamical mechanism, such as retroactions, contributes to determine them; they arise purely from the structure of the measure on the boundary conditions of the universe. Certainly, no experimenter can intentionally prepare in the laboratory two particles distributed according to $\eta_-$ or according to $\eta_\pm$; however, as we have already noted, determining the boundary conditions of the universe cannot be likened to preparing a physical system for a laboratory experiment.

Let us summarize the concept of preinteractive correlations with the following statement:
\begin{quote}
{\it Preinteractive correlations occur when the statistical distribution of the particles in a system (also) depends on the interactions that the system will experience in the future; preinteractive correlations do not have a dynamical cause but arise exclusively from the structure of the measure on the boundary conditions of the universe.}
\end{quote}

It is easy to realize that if one allows the possibility of preinteractive correlations the EPR paradox can no longer be demonstrated. Consider the classical proof of the paradox \cite{bell2}, slightly modified in order to highlight the point where the existence of preinteractive correlations is implicitly excluded. Four identically prepared sources consist of a large number of atomic systems. The atomic systems emitted by the four sources undergo four different types of measurement. The state of each atomic system is completely represented by a hidden variable $\lambda$, which univocally determines the result of any measurement performed on the atomic system. The statistical distribution of $\lambda$ in the four sources is described by four normalized densities $\rho_i(\lambda)$, where $i=1,...,4$ indicate the source. Since the four sources have undergone the same preparation process, the densities $\rho_i(\lambda)$ are equal. Bell's inequality is demonstrated starting from these hypotheses.

It is immediately evident that assuming the equality of the four densities $\rho_i(\lambda)$ means excluding the existence of preinteractive correlations. This assumption is in fact based on the hypothesis that the densities $\rho_i(\lambda)$ depend on the source preparation phase - i.e., on interactions in the past -, not on the measurement phase - the interactions in the future. It is trivial to show that, if this assumption is removed, allowing the densities $\rho_i(\lambda)$ to be different for different kinds of measurement, Bell's inequality can no longer be demonstrated.

This workaround for the EPR paradox, at least from the formal point of view, is known (Bell\cite{bell2}, Price, Szab\'{o}\cite{szabo1}\cite{szabo2}, Durt\cite{durst1}\cite{durst2}\cite{durst2}). Many authors, however, find it so absurd that they do not even take it into consideration and refer to it with terms such as ``global conspiracy of probabilities", ``fatalism", ``superdeterminism", ``theory of prearranged harmony", and so on. The paradoxical aspect of this workaround arises both from the presence of a counterintuitive phenomenon such as preinteractive correlations and from the explicit need to include the experimenters in the overall deterministic description of the system, thus denying them the possibility to freely choose the measurement to be performed on the atomic systems. We believe that showing the link between preinteractive correlations and the choice of the F-type boundary conditions of the universe may help to make this workaround more acceptable; as regards determinism, in section 13 we will show that the deterministic character of the quantum-classical theory is probably not one of its fundamental aspects, since it disappears when dynamical phenomena not predicted by a hamiltonian system of particles, such as particle decay, occur.

Although we have indicated this possible workaround for the EPR paradox, we must add that the quantum-classical theory, at least in its present formulation, is unable to explain the paradox completely. The theory cannot represent particles with spin, and thus cannot provide an explicit model for the EPR experiment; obtaining a model of this kind requires extending the quantum-classical theory. Such an extension may entail the appearance of {\it non local correlations}, having the same non-dynamical nature as preinteractive correlations, and the explanation of the paradox may be based more correctly on non-local correlations rather than on preinteractive correlations.

Let us now examine in general the role of preinteractive and postinteractive correlations in determining the characteristics of the evolution of our universe. Price considers preinteractive independence - i.e., the lack of preinteractive correlations - on two different levels: the macroscopic level and the microscopic level. At the macroscopic level, preinteractive independence is linked to the second law of thermodynamics and to the arrow of time. It is well known that Boltzmann's H theorem is based on the assumption that the momenta of interacting microsystems are not statistically correlated before the collision. More intuitively, preinteractive independence, at the macroscopic level, is found in the apparent time-asymmetry of evolution, due to which one can observe, for example, stones falling into the water and making ripples which diverge from the center but never observes ripples converging to the center and ejecting stones from the water. One can therefore state that at the macroscopic level preinteractive independence is an unquestionable and objective evidence.

On the contrary, at the microscopic level preinteractive independence is not that evident. Indeed, as Price noted, the implicit and unconditional assumption of preinteractive independence is probably the source of many of the paradoxical aspects of quantum mechanics, such as the EPR paradox or delayed choice experiments \cite{wheeler}, which we will not discuss here. We encountered another example of preinteractive correlations linked to quantum mechanics while studying the scattering process, where we have seen that if one admits the presence of preinteractive correlations it is possible to reproduce the quantum mechanical cross-section in a classical dynamical scheme.

We have seen that preinteractive and postinteractive correlations are determined by the structure of the measure on the boundary conditions of the universe. The actual measure on the boundary conditions of our universe should therefore predict a high prevalence of postinteractive correlations, in order to explain the second principle of thermodynamics and the arrow of time, and a small portion of preinteractive correlations, linked to quantum phenomena. Consider the two densities $\rho_{FC}$ and $\rho_{FQ}$ studied in section 7 from this point of view. Classical density corresponds to initial conditions of molecular chaos, i.e. to initial particle momenta which are completely independent and uncorrelated; it is therefore reasonable to claim that it determines an evolution which is completely dominated by postinteractive correlations. Equation (\ref{9}) shows that the quantum density $\rho_{FQ}$ (in semiclassical approximation) differs from the classical density basically in the interference term $\rho_{FI}$; it is therefore reasonable to hold this term responsible for preinteractive correlations of  the quantum type.

\section{Determinism, indeterminism and the least action principle.}

The quantum-classical theory is based on classical mechanics, which is a deterministic theory, i.e. a theory in which initial conditions plus law of motion determine the path univocally. Nonetheless, determinism is probably not a fundamental aspect of the quantum-classical theory, since it disappears when the theory is extended to encompass more complex phenomena, such as particle decays or electron-photon interactions, predicted by quantum field theory. This happens because in a system in which the number of particles is conserved, the least action principle and the type $F$ boundary conditions are more or less equivalent to Hamilton's equations plus type $I$ boundary conditions, whereas in a system which allows particle decays or particle absorption this is no longer true. Consider for example the graph of figure 3, which plots the decay of a particle of mass $m_1$ into two particles of masses $m_2$ and $m_3$.

\begin{center}
\unitlength=1mm
\begin{picture}(90,45)
\put(2,25){\vector(1,0){30}}
\put(32,25){\vector(4,1){50}}
\put(32,25){\vector(2,-1){30}}

\put(2,25){\circle*{1}}

\put(-3,18){\makebox(10,6){$({\bf x}_1,t_I)$}}
\put(12,25){\makebox(10,6){1}}
\put(25,18){\makebox(10,6){$({\bf x},t)$}}
\put(57,32){\makebox(10,6){2}}
\put(65,7){\makebox(10,6){$({\bf x}_3,t_F)$}}
\put(42,12){\makebox(10,6){3}}
\put(85,35){\makebox(10,6){$({\bf x}_2,t_F)$}}

\end{picture}

Fig. 3
\end{center}

The type $I$ boundary conditions are the position ${\bf x}_1$ and the momentum ${\bf p}_1$ of particle 1 at the time $t_I$, whilst the type $F$ boundary conditions are the position ${\bf x}_1$ of particle 1 at the time $t_I$ and the positions ${\bf x}_2$, ${\bf x}_3$ of particles 2 and 3 at the time $t_F$ \footnote{In this case, boundary conditions should more correctly include the masses of the decay particles.}. It is straightforward to notice that Hamilton's equations plus the type $I$ boundary conditions are unable to determine the decay coordinates $(x,t)$ or the decay particle momenta; however, this is possible if one uses the least action principle together with type $F$ conditions, as we will now show with simple calculations. Consider, for the sake of simplicity, the low-velocity approximation for a one-particle relativistic lagrangian:
$$
L=-mc^2+\frac{1}{2}m{\bf v}^2-V.
$$
For a vanishing potential, the least action path connecting two points is a straight line covered at constant velocity. Therefore we can write the action $S$ of the process shown in figure 3 in the following way:
\begin{eqnarray}
S & = & -m_1c^2(t-t_I)+ \frac{1}{2}m_1\frac{({\bf x}-{\bf x}_1)^2}{t-t_I} 
-m_2c^2(t_F-t)+ \frac{1}{2}m_2\frac{({\bf x}_2-{\bf x})^2}{t_F-t} \\
& & -m_3c^2(t_F-t)+ \frac{1}{2}m_3\frac{({\bf x}_3-{\bf x})^2}{t_F-t}.\nonumber
\label{39}
\end{eqnarray}
The action is minimized by equating to 0 its derivatives with respect to ${\bf x}$ and $t$, thus:
\begin{equation}
m_1\frac{{\bf x}-{\bf x}_1}{t-t_I}
-m_2\frac{{\bf x}_2-{\bf x}}{t_F-t }
-m_3\frac{{\bf x}_3-{\bf x}}{t_F-t }=0,
\label{40}
\end{equation}
$$
-m_1c^2-\frac{1}{2}m_1\frac{({\bf x}-{\bf x}_1)^2}{(t-t_I)^2}+
m_2c^2+\frac{1}{2}m_2\frac{({\bf x}_2-{\bf x})^2}{(t_F-t)^2}+
m_3c^2+\frac{1}{2}m_3\frac{({\bf x}_3-{\bf x})^2}{(t_F-t)^2}=0.
$$
i.e. the equations of momentum and energy conservation. Equations (\ref{40}) form a system of four equations that allow to compute the four unknown variables $(x,t)$. The equation of momentum conservation leads to:
\begin{equation}
\frac{m_1}{t-t_I}({\bf x}-{\bf x}_1)=\frac{m_2+m_3}{t_F-t}({\bf x}_{CM}-{\bf x}),
\label{41}
\end{equation}
where
$$
{\bf x}_{CM}=\frac{m_2{\bf x}_2+m_3{\bf x}_3}{m_2+m_3}.
$$
Without loss of generality, we can work in a reference frame where ${\bf x}_1={\bf x}_{CM}=0$. Equation (\ref{41}) also allows to find that ${\bf x}=0$\footnote{This holds for the solution with $t_I\leq t\leq t_F$ which we are seeking.}, and the equation of energy conservation allows one to obtain:
$$
t=t_F-\sqrt{\frac{m_2{\bf x}_2^2+m_3{\bf x}_3^2}{2(m_1-m_2-m_3)c^2}},
$$
from which the momenta of particles 1 and 2 can be computed trivially.

\vspace{5mm}
In conclusion, one can say that the least action principle, together with the type $F$ boundary conditions, allows to determine paths which Hamilton's equations plus the initial conditions are unable to determine; this means that these paths are not deterministic.

\section{Conclusion}

Let us summarize the main features of the quantum-classical theory.

It consists of two distinct laws, one of probabilistic nature, the other of dynamical nature:
\begin{itemize}
\item[1.]the probabilistic law is the quantum measure on the boundary conditions of the universe, which gives rise to the statistical regularities of evolution.
\item[2.]the dynamical law is the least action principle, which determines the path.
\end{itemize}

The aspects of evolution governed by these two laws are as distinct as the principles from which they are derived: the quantum measure is derived from the sum of the amplitudes of all the paths having the same final position (i.e. with the same p-limit), whilst the least action principle selects, among all the paths, the one that minimizes the action.

The fact of considering the least action principle as a dynamical law instead of Hamilton's equations is fundamental for two reasons: first of all, it provides the correct type of boundary conditions to which the quantum measure is to be applied; moreover, together with the corresponding boundary conditions, it allows to define a class of paths which is broader than the one defined by Hamilton's equations with initial conditions; this extension is necessary in order to represent phenomena such as particle decay. Therefore, determinism is not a fundamental feature of the least action principle and thus of the quantum-classical theory.

We have shown the analogies and differences between the quantum-classical theory and Bohm's causal theory: in both theories there is a set of paths and a statistical distribution. The difference between the two theories, as shown in section 11, lies in the fact that whilst Bohm's theory requires the statistical distribution $\rho(x,t)$ of the paths and the square of the wave function $|\psi(x,t)|^2$ to be rigorously equal for every $x$ and $t$, the quantum-classical theory requires them to be only ``macroscopically" equal. This weaker constraint arises from the consideration that our perception of macroscopic phenomena is always mediated by macroscopic instruments; therefore two theories are equivalent if they describe macroscopic instruments recording the same results in the measurements. By dropping the requirement of precise equality between $\rho(x,t)$ and $|\psi(x,t)|^2$, the need for a quantum potential in the Hamiltonian disappears, together with the need to attribute a real physical meaning to the wave function. Therefore, in the quantum-classical theory the problem of wave function collapse is not posed at all, since the wave function does not exist.

The conceptually most important proposal of the quantum-classical theory is certainly to use the conditions associated with the least action principle, rather than the initial conditions, as boundary conditions for the universe. We note that this proposal is supported by the observation that defining the boundary conditions of the universe is fundamentally different from preparing a system for a laboratory experiment. We have seen that this viewpoint can lead to apparently paradoxical phenomena, such as preinteractive correlations, and that when preinteractive correlations are present, Bell's inequality can no longer be demonstrated. Therefore, whilst on the one hand this proposal certainly has paradoxical, or at least conceptually unusual, aspects (which in any case correspond to the paradoxical aspects that quantum mechanics itself exhibits), on the other hand it has the advantage of being based on a universal physical principle such as the least action principle.

The boundary conditions associated with the least action principle do not identify the path univocally. Therefore the different results of a measurement, which obviously correspond to different paths, might have the same boundary conditions. In order to preserve the internal consistency of the theory, however, the conjecture has been made that the paths that correspond to different results of a measurement, and more generally paths that differ macroscopically from each other, have different boundary conditions. This conjecture allows every measurement to be traced back to a position measurement made at the final instant of the universe.

This paper has examined two physical examples in the light of the quantum-classical theory: the two-slit experiment and the scattering process. However, a rigorous demonstration that this theory ``works" in any situation has not been provided. Such a demonstration, if possible, probably entails specifying and thoroughly verifying the ``macroscopic" equality between the quantities $\rho'_{FQ}(x,t)$ and $|\psi(x,t)|^2$. It may also be possible to provide a ``negative" demonstration, i.e. to provide a precise example of an experimental situation that the quantum-classical theory cannot explain; obviously, in this case the theory would have to be abandoned. A more precise answer to this question can only be arise from further studies and verifications of the theory.

\section{Appendix.}

This appendix is a more thorough study of the statistical regularities of the Bernoulli application introduced in section 3.

The proof that Lebesgue's measure on the interval $[0,1)$ determines for $\alpha$ the value 1/2 can be found in many texts of probability theory, such as \cite{probability}\footnote{In this text, the demonstration refers to the decimal expansion of a number. However, it can be also applied to its binary expansion.}.

We show that any statistical regularity in the binary expansion of the initial condition can be originated by a suitable measure on the interval $[0,1)$. For this purpose we will use notions, such as {\it cylindrical sets} and {\it marginal distributions}, which belong to stochastic process theory. Let $Z\subseteq [0,1)$ be the set of numbers of the form $m/2^n$, with $n$ and $m$ positive integers. The numbers of $[0,1)\setminus Z$ have a single binary expansion, whilst the numbers of $Z$ have two binary expansions; by convention, we choose the expansion that ends with all zeroes. We indicate with $\{0,1\}^\infty$ the set of all the possible sequences of elements 0 and 1; with $\xi$ the map $\xi:[0,1)\rightarrow \{0,1\}^\infty$ which associates with every $a\in [0,1)$ its binary expansion, and with $X$ the range of the map $\xi$, so that $\xi$ is bijective between $[0,1)$ and $X$. A cylindrical subset of $X$ is a set of the type $\{\{a_i\}\in X:a_{i_1}=c_1,...,a_{i_{n}}=c_n\}$ where $i_1<...<i_n$ is a generic finite sequence of integers, and $c_1,..., c_n\in {0,1}$. We indicate with ${\cal C}_0$ the class of cylindrical sets of $X$; it can be shown that ${\cal C}_0$ is an algebra; we also indicate with ${\cal C}$ the $\sigma$-algebra of $X$ generated by ${\cal C}_0$, i.e. the smallest $\sigma$-algebra that contains ${\cal C}_0$.

We prove that ${\cal B}=\xi^{-1}({\cal C})$, where ${\cal B}$ is Borel's $\sigma$-algebra on $[0,1)$. If $C\in {\cal C}_0$, then $\xi^{-1}(C)$ is the union of a finite number of semiopen intervals $[r,s)$, where $r,s\in Z$, and ${\cal B}_0\equiv \xi^{-1}({\cal C}_0)$ is the algebra of $[0,1)$ constituted by the finite unions of such intervals. Let $\sigma({\cal B}_0)$ the $\sigma$-algebra generated by ${\cal B}_0$; since ${\cal B}$ is generated for example by the class of semiopen intervals, which contains ${\cal B}_0$, we find that $\sigma({\cal B}_0)\subseteq {\cal B}$. Since $Z$ is dense in $[0,1)$, an open interval $(a,b)\subseteq [0,1)$ can always be expressed as countable union of intervals of ${\cal B}_0$; however, since ${\cal B}$ is generated by the open intervals as well, we find that ${\cal B}\subseteq\sigma({\cal B}_0)$, and therefore ${\cal B}=\sigma({\cal B}_0)$. This fact, the fact that ${\cal B}_0=\xi^{-1}({\cal C}_0)$ and the fact that $\xi$ is bijective between $[0,1)$ and $X$, lead to the thesis.

A generic statistical regularity on the set $X$ corresponds to a family of marginal distributions $\nu_0$ on ${\cal C}_0$, which by virtue of the measure extension theorem can be extended univocally to a measure $\nu$ on ${\cal C}$; finally, the measure $\nu$, by means of the map $\xi$, induces a measure $\mu$ on ${\cal B}$. This means that every statistical regularity in the binary expansion of a number of the interval $[0,1)$ corresponds to a suitable measure on $[0,1)$.

Consider for instance the family of marginal distributions $\nu_0$ defined by letting $\nu_0(C)\equiv\alpha^p(1-\alpha)^q$, where ${\alpha}$ is a fixed number between 0 and 1, $C=\{\{a_i\}\in X:a_{i_1}=c_1,...,a_{i_{n}}=c_n\}$, $p=n-\sum c_i$ and $q=\sum c_i$; this definition is equivalent to the attribution of a probability $\alpha$ to digit 0 and $(1-\alpha)$ to digit 1 of the binary expansion. This statistical regularity is matched by a measure $\mu$ on $[0,1)$, which is produced in the manner shown earlier. This measure is generally singular with respect to Lebesgue's measure. From the law of large numbers one obtains that the measure $\nu$ of the sequences (and therefore the measure $\mu$ of the numbers) for which the quantity $n_0/n$ does not tend to $\alpha$ is null.


\end{document}